# An Adaptive Neuro-Fuzzy Blockchain–AI Framework for Secure and Intelligent FinTech Transactions

Gunjan Mishra[a*], Yash Mishra[b]

[a] Retail Banking, Debit card, Capital One, McLean, Virginia, USA, gunjan.mishra@capitalone.com
[b] JK Lakshmipat University, India, yashmishra@jklu.edu.in

**Abstract**
Financial systems have a growing reliance on computer-based and distributed systems, making FinTech systems vulnerable to advanced and quickly emerging cyber-criminal threats. Traditional security systems and fixed machine learning systems cannot identify more intricate fraud schemes whilst also addressing real-time performance and trust demands. This paper presented an Adaptive Neuro-Fuzzy Blockchain-AI Framework (ANFB-AI) to achieve security in FinTech transactions by detecting threats using intelligent and decentralized algorithms. The framework combines both an immutable, transparent and tamper resistant layer of a permissioned blockchain to maintain the immutability, transparency and resistance to tampering of transactions, and an adaptive neuro-fuzzy learning model to learn the presence of uncertainty and behavioural drift in fraud activities. An explicit mathematical model is created to explain the transaction integrity, adaptive threat classification, and unified risk-based decision-making. The proposed framework uses Proof-of-Authority consensus to overcome low-latency validation of transactions and scalable real-time financial services. Massive simulations are performed in normal, moderate, and high-fraud conditions with the use of realistic financial and cryptocurrency transactions. The experimental evidence proves that ANFB-AI is always more accurate and precise than recent state-of-the-art algorithms and costs much less in terms of transaction confirmation time, propagation delay of blocks and end-to-end latency. ANFB-AI performance supports the appropriateness of adaptive neuro-fuzzy intelligence to blockchain-based FinTech security.

## 1. Introduction

The emergence of digital financial services has changed the face of conventional banking, allowing financial operations to be performed quicker, more conveniently, and more effectively on a global scale. Nevertheless, it is the same digital evolution that poses increased cybersecurity threats, such as fraud, identity theft, and unauthorized access to sensitive financial information [1]. As a result, financial institutions and FinTech firms are in the process of finding technological answers to protect digital dealings and improve efficiency in their operations. The blockchain technology has turned out to be a promising solution to counter the cybersecurity threats in financial transactions. Using its decentralized system of ledger keeping, the immutable record keeping and the inference by consensus, blockchain mitigates the chances of fraud or unauthorised operations in banking systems [2]. Studies have proven that blockchain implementation into financial activities does not only enhance the levels of security but also facilitates the verification of transactions in different institutions [3].

Blockchain applications have demonstrated potential in trade finance and banking security by preventing fraud, improving transparency, and supporting smart contracts to enforce compliance automatically [4]. Furthermore, the intersection of blockchain and FinTech novelty can be used to establish dynamic systems that can effectively react to any changing cyber threat [5]. The developments are essential because the number and complexity of financial cyber-attacks are on the rise across the world. AI-based methods and deep learning are also useful complement to blockchain as they offer predictive risk analysis and anomaly detection in banking activities. Hybridization systems AI + blockchain enable financial institutions to predict areas of their vulnerability, suspicious transactions, and anticipate security breaches in real time [6]. Such integration is the foundation of intelligent and automatic risk management frameworks in the current banking. The most recent research has been devoted to blockchain-based frameworks specifically oriented to FinTech banking industries. Indicatively, optimized blockchain deployment in banking networks has been predicted in adaptive neuro-fuzzy-based K-nearest neighbours' algorithms, and these algorithms have shown enhanced reliability of transactions and a robust system [11]. The conceptual block diagram of FinTech in banking presented in Figure 1 demonstrates customer engagement, edge processor, blockchain ledger, AI-based monitoring, and central financial control. The systematic literature reviews prove that blockchain in financial institutions improves not only the security but operational transparency and auditability. Banks that use blockchain-based solutions note that the time taken to settle transactions and compliance levels have decreased [7]. Moreover, blockchain supports the cross-border payments, verification of digital identities, and the safe management of contracts, which is vital to the international financial activities [8].

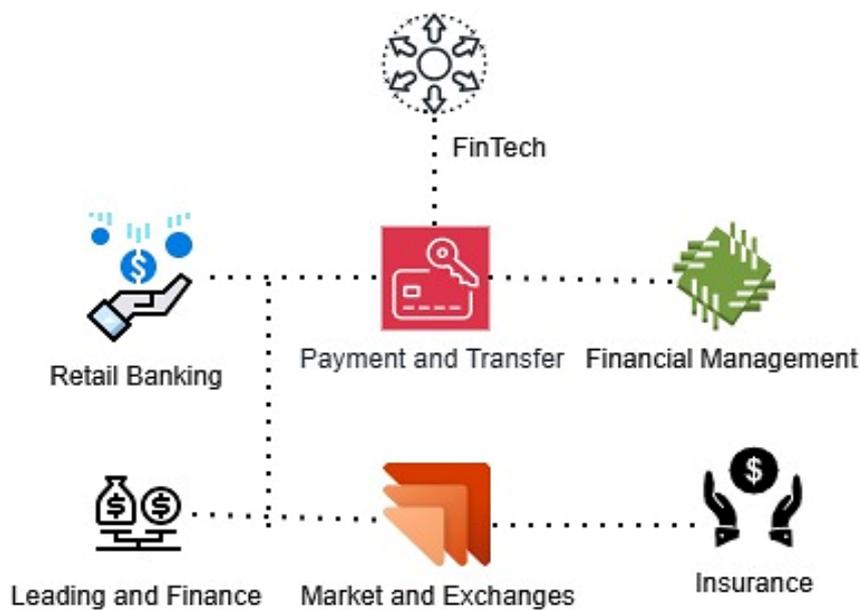

**Figure 1.** General representation of Financial Technology (FinTech) in banking operations.

Bibliometric analyses suggest a growing research momentum around blockchain and cybersecurity applications to banking and FinTech with the latest research highlighting how the distributed ledger technology, cloud computing, and AI can be converged to provide secure and scalable financial networks [9], [10]. These works confirm the necessity of complex structures that combine various technologies to tackle the issues of operational efficiency and security challenges at the same time. To conclude, it is becoming more accepted that blockchain technology and blockchain technology in conjunction with AI and predictive analytics can be the foundation of the next generation FinTech banking frameworks. These systems do not only protect against cybersecurity threats, but also make it possible to have real-time monitoring, predictive risk assessment and automated mitigation plans [11]. This paper presented a detailed blockchain-based FinTech system that uses these technologies to support security, efficiency, and resiliency of the financial institutions. This paper makes three primary contributions at the intersection of FinTech, blockchain, and cybersecurity in banking:
   1. **FinTech Framework based on blockchain:** We introduce a novel framework combining blockchain, AI, and adaptive neuro-fuzzy algorithms to improve cybersecurity, integrity of transactions, and risk assessment in real-time in digital banking.

2. **AI Adaptive AI-driven Risk Management:** The model provided a AI-driven predictive analytics and anomaly detection to track financial transactions in real-time, identify possible fraud count, and offer proactive remediation measures.
3. **Empirical validation and efficiency:** we used synthetic and benchmarked banking data to show that the proposed framework enhances security, transparency of transactions and efficiency in operations in comparison to traditional FinTech systems.

The rest of this paper is organized in the following way. Section 2 is a literature review of related work on blockchain applications, FinTech cybersecurity, and AI-enabled financial systems. Section 3 discusses the proposed blockchain-based FinTech framework, architectural design and algorithmic components. Section 4 details the experimental setup, evaluation metrics, and presents results and discusses the framework's effectiveness in enhancing cybersecurity and operational efficiency. Finally, Section 5 concludes the paper and outlines future research directions.

## 2 Related work

The financial services are quickly becoming digitalized, which has created both greater opportunities and major cybersecurity challenges faced by banks and FinTech institutions. Conventional rule-based fraud detection systems cannot cope with the rising complexity and volume of digital transactions, which tends to lead to high false-positives and slow response rates. Machine learning (ML) and deep learning (DL) methods have become effective instruments to infer abnormalities and fraud in financial systems. ML algorithms that are supervised, including Random Forests, Support Vector Machines (SVM), and gradient boosters like XGBoost have demonstrated excellent performance in structured transaction data, which has excellent precision and recall rates [12]–[14].

DL models, such as Long Short-Term Memory (LSTM) networks, autoencoders, and transformer-based designs have been shown to be able to recognize sequential and contextual anomalies and capture complex patterns that traditional ML approaches can often overlook [15], [16]. Although centralized ML and DL solutions are predictive, they put sensitive financial information at risk of privacy breaches and means of single-point failures, which underscores the necessity of decentralized and privacy-conscious solutions. The blockchain technology has been internationally deemed as one of the potential solutions to financial transaction integrity, immutability and transparency. With the help of decentralized ledgers and consensus mechanisms, blockchain can decrease the possibility of fraud and unauthorized alterations [17]–[19].

In trade finance and banking, smart contracts based on blockchain have been applied to automate compliance, achieve greater transparency in cross-border trade, and generate records that are immutable and can be used in an audit [18], [20]. Several studies have considered the combination of ML models and blockchain systems to store the results of anomaly detection in an irrevocable way and allow decentralized monitoring [19], [21]. Nevertheless, the majority of the blockchain applications are aimed at post-hoc verification and not real-time threat detection, which restricts their performance in addressing high-speed cyber threats in the FinTech setting.

The idea of federated learning (FL) has become one of the possible solutions to the problem of privacy in collaborative training of AI models. FL prevents access to sensitive transaction data and enhances the predictive accuracy by enabling several financial institutions to train models locally and send only encrypted updates [22], [23].

The addition of FL to blockchain improves the level of trustworthiness even more since this solution offers verifiable and tamper-resistant updates on the model [24], [25]. Although FL allows collaborating privacy-preserving, current blockchain-FL schemes mostly ignore dynamic threat detection and dynamically adjusting risk evaluation, which is vital to proactive cybersecurity during financial transactions. Additionally, little studies have focused on the implementation of neuro-fuzzy-based AI models in this regard that can provide interpretable and adjustable decision-making processes on complex transactional patterns. Hybrid AI-blockchain systems have been considered as one of the solutions of filling the gap between predictive intelligence and transactional security. In these frameworks, AI models of anomaly detection are combined with blockchain-ledger to provide auditability and tamper-proof logging [20], [26].

Big data analytics allows larger volumes of transactions to be processed in real-time, whereas explainable AI approaches allow transactions to be understandable to meet compliance and audit needs [21], [27]. Still, most of the current hybrid architecture pay attention to either operational security or predictive analytics as a standalone solution, and they hardly implement adaptive AI, blockchain, or federated learning, active-cybersecurity framework to FinTech banking. This deficiency of built-in structures that can detect anomalies in real time, forecast risks as well as the integrity of transactions illustrates a research gap. The recent literature reviews were done separately on blockchain applications, AI-based fraud detection, and federated learning approaches [28]–[30].

To illustrate, research in fraud detection using blockchain has shown that it will enhance transparency and auditability, especially in credit card and cryptocurrency transactions [28], [29].

It has been demonstrated that federated learning methods can enhance the accuracy of models and maintain privacy in distributed nodes [30]. Nonetheless, the void exists in any detailed models combining blockchain, AI, and FL to facilitate the real-time detection of threats, adaptive risk control, and operational efficiency all at once. In addition, limited research has been conducted on the adaptive neuro-fuzzy algorithms application in this area, which are specifically appropriate in managing uncertainty and changing trends in financial transactions [31].

**Table 1.** Literature review comparison.

| Ref | Study Focus | Techniques Used | Key Findings / Gaps |
| --- | --- | --- | --- |
| [12] | Credit card fraud detection | Random Forest, SVM | High precision and recall, centralized, privacy risk |
| [13] | Financial transaction anomaly detection | XGBoost | Accurate prediction, lacks integration with blockchain |
| [14] | Cross-bank fraud detection | Gradient Boosting, Logistic Regression | Good predictive performance, no decentralized approach |
| [15] | Sequential fraud detection | LSTM, Autoencoders | Captures temporal patterns, centralized, limited privacy |
| [16] | Multi-merchant fraud detection | Transformers | Contextual anomaly detection, computationally intensive |
| [17] | Blockchain in trade finance | Blockchain + Smart Contracts | Ensures integrity, lacks AI-based anomaly detection |
| [18] | Cross-border payment security | Blockchain | Transparent and immutable, limited real-time analysis |
| [19] | Blockchain-ML integration | Blockchain + ML | Post-hoc fraud verification, no adaptive risk management |
| [20] | Hybrid AI-blockchain frameworks | AI + Blockchain | Improved auditability, lacks integrated FL and adaptive models |
| [21] | Explainable AI in finance | XAI | Interpretability for compliance, limited integration with blockchain |
| [22] | Federated learning for banks | FL | Preserves privacy, no real-time threat detection |
| [23] | Privacy-preserving AI | FL | Collaborative learning without data sharing, lacks adaptive models |
| [24] | Blockchain + FL | Blockchain + FL | Tamper-resistant model updates, limited threat analysis |
| [25] | Distributed fraud detection | FL | Improves accuracy and privacy, not real-time |
| [26] | AI-blockchain for transaction monitoring | AI + Blockchain | Detection accuracy improved, operational efficiency not evaluated |
| [27] | Big data analytics for fraud | Big Data + AI | Processes high-volume transactions, lacks blockchain integration |
| [28] | Systematic review: blockchain fraud detection | SLR | Improved transparency, gaps in real-time detection |
| [29] | Cryptocurrency fraud detection | Blockchain | Auditability enhanced, limited dynamic risk assessment |
| [30] | Federated ML for financial systems | FL | Privacy-preserving, lacks adaptive AI and blockchain integration |

| [31] | Neuro-fuzzy adaptive models | Neuro-fuzzy KNN | Handles uncertainty, rarely applied in integrated blockchain-FL systems |

Overall, the literature indicates that although considerable progress has been achieved around ML/DL, blockchain, and federated learning to tackle financial cybersecurity, some of the most critical issues are yet to be addressed. The sensitive data of the centralized ML expose to a risk of privacy, the blockchain implementations in many cases cannot provide real-time detection, and the available FL frameworks lack full adaptive AI models to provide dynamic risks. Such research gaps directly drive the contribution of the given paper: to design a blockchain-based FinTech framework, including AI-based threat identification, adaptive neuro-fuzzy risk control, and operational efficiency analysis based on empirical confirmation provided on benchmark examples.

Although the field of blockchain, AI, and machine learning around financial fraud detection has evolved rapidly, current research indicates that the practice has several significant vulnerabilities that reduce the effectiveness and feasibility of FinTech cybersecurity systems. Most machine learning methods, though with high detection rates, use centralized models that are privacy threatening and have the weakness of single point failures [12], [13] as exhibited in table 1. Federated learning reduces privacy issues but usually experiences non-IID information disseminations and uneven model conduct in dispersed PCs [13], [14]. On the same note, blockchain solutions make systems more invoke the property of immutability and auditability, but their incorporation with AI-driven anomaly detection is limited, which leads to systems that are either security-centric or predictive intelligence-centric, but lack decentralized trust [17], [25], [26]. Also, there is a dearth of research offering an integrated system integrating blockchain, adaptive AI, and neuro-fuzzy algorithms to manage real-time threat detection and still be operationally efficient [11], [24], [27]. This was the gap that indicates that no end-to-end frameworks were developed that could handle privacy, scalability, predictive accuracy, and real-time monitoring concurrently in FinTech systems. In the current research, we attempt to fill these gaps by developing a blockchain-based cybersecurity architecture incorporating AI-based threat recognition and adaptive neuro-fuzzy controls, which could ensure the integrity of transactions, improved predictive risk management, and operational resilience in online financial systems [34].

## 3  Proposed Work

This section the proposed blockchain-based cybersecurity framework aiming at securing FinTech transactions by integrating distributed ledger technology and AI-based threat detection systems are introduced. The framework is inspired by the shortcomings that have been identified in the literature especially the absence of integrated architectures that can address transaction integrity, real-time threat detection, privacy preservation, and scalability in financial systems all at the same time. The proposed framework will offer a resilient and adaptive security infrastructure to the banking environment of the present day by integrating the decentralized trust framework of blockchain with intelligent analytics. We have presented the detailed system architecture of the proposed framework and the proposed framework.

## 3.1 System Architecture

The proposed framework has a system architecture made up of connected functional layers that jointly safeguard financial transactions lifecycle. The user interaction layer is used when the customer engages in financial transactions using mobile banking applications, digital wallets or online banking portals. These are transactions that are pre-authenticated through basic authentication and encryption systems and then transmitted to be processed further. The metadata of transactions such as timestamps, transaction value, location and behavioural features are gathered and processed at the level of transaction processing and data acquisition. The sensitive customer data are made anonymous to respect privacy prior to being sent to the analysis pipeline. This layer is where the traditional banking systems and the intelligent security modules interface. The security framework is based on the blockchain layer. The transactions are signed and stored as hash functions cryptographically and authenticated by consensus mechanisms like Practical Byzantine Fault Tolerance (PBFT) or Proof of Authority (PoA) that perimeter permissioned banking networks. Smart contracts are utilized to implement security policy, automated compliance checks and to issue alerts when suspicious behaviour is identified. The layer of AI-supported threat detection follows the streams of transaction logs present on the blockchain. Machine learning algorithms study historical and real-time transaction patterns to detect any anomalies which could represent a fraud, ransomware or identity theft attempt.

The adaptive neuro-fuzzy inference systems also elevate interpretability and decision making as they integrate the fuzzy logic and learning, and so the framework can process uncertainties and inaccurate financial data. Last but not the least, the response and governance layer orchestrates automated responses and human-in-the-loop responses. Upon detection of a threat, the system can freeze transactions temporarily, alert security analysts, and record the incidents in an irrevocably read-only way to the blockchain to be audited and compliant with regulations. This layer helps to make the security actions transparent, traceable, and responsible.

### 3.1.1 AI-Driven Threat Detection and Learning Mechanism

The smartness of the suggested framework is its AI-based threat detection system. The framework uses both supervised and unsupervised learning models to model transactional behaviours as opposed to rule-based security systems. The training of supervised models is done using a labelled dataset of frauds that help identify known patterns of attacks, whereas the training of unsupervised models is done using new anomalies that have never been seen before. The framework can be used to overcome privacy issues presented by centralized learning by facilitating decentralized model updates, that is, model parameters are only shared, not financial data. Adaptive neuro-fuzzy models integration facilitates explainable decision-making, which is important in a banking system in terms of regulatory compliance. This mixed system enables the financial institutions to strike a balance between accuracy of detection and transparency and trust.

### 3.1.2 Blockchain Integration and Security Enforcement

Blockchain The blockchain is used to make sure transactions are immutable, data is intact, and non-repudiable throughout the financial network. Cryptographically signed transactions can be verified to store the records with previous transactions, and thus unauthorized modifications cannot be computedally feasible. Smart contracts serve as independent security intermediaries that impose established rules, e.g. transaction limits, multi-factor checks and compliance restrictions. The framework provides the opportunity to reinforce security policies in real time using AI-generated risk scores implemented into the smart contracts. Risky transactions could be flagged or held in a pending position to be checked manually by a customer, but low-risk transactions would not be hampered. Such selective enforcement ensures that the overheads of operation are minimized yet the security levels are retained at a high level.

### 3.1.3 Operational Workflow

The working process starts when a client opens a financial operation with the help of a FinTech interface. It encrypts the transaction and anonymizes it and sends the encrypted transaction to the processing layer where it is subjected to initial checks. The network is then sent to the blockchain network to be validated and documented. At the same time, the AI-based module analyses the features of transactions to calculate a risk score. In case of legitimate transaction, the transaction is complete and stored permanently on the blockchain. When a suspicious activity is detected, the framework reacts to security mechanisms, entries the event, and corrects the learning models to enhance detection in future. This feedback mechanism will ensure that the performance of security is continuously improved as time goes by.

## 3.2 Proposed Adaptive Neuro-Fuzzy Blockchain-AI Framework (ANFB-AI)

The proposed an Adaptive Neuro-Fuzzy Blockchain-AI Framework (ANFB-AI) is based on four principal concepts, which include decentralization, transparency, intelligence, and adaptability. The blockchain technology attains this through decentralization, removes single points of failure and provides tamper-resistant records of transactions. Transparency and auditability are offered through irreversible ledges that enable financial institutions and regulators to confirm transactions without revealing sensitive customer information. Intelligence is also presented by AI-based threat detection models that are able to detect anomalous patterns in real time, and adaptability is backed by continuous learning new models to overcome the current cyber threats. The suggested Blockchain-AI FinTech architecture presented in figure 2, which makes the design options strictly specified and analytically solvable. The modelling is intended to reflect three central system goals: (i) to provide secure and immutable transaction processing using blockchain technology, (ii) to offer adaptive and data-driven fraud and threat detection under uncertainty, and (iii) to provide scalable real-time decision-making that is suited to the high throughput FinTech setting. Incorporating computational intelligence, in addition to the decentralized trust, through the abstracted FinTech ecosystem as a distributed cyberphysical system, the proposed model will mitigate the security and performance limitations of the contemporary digital financial systems.

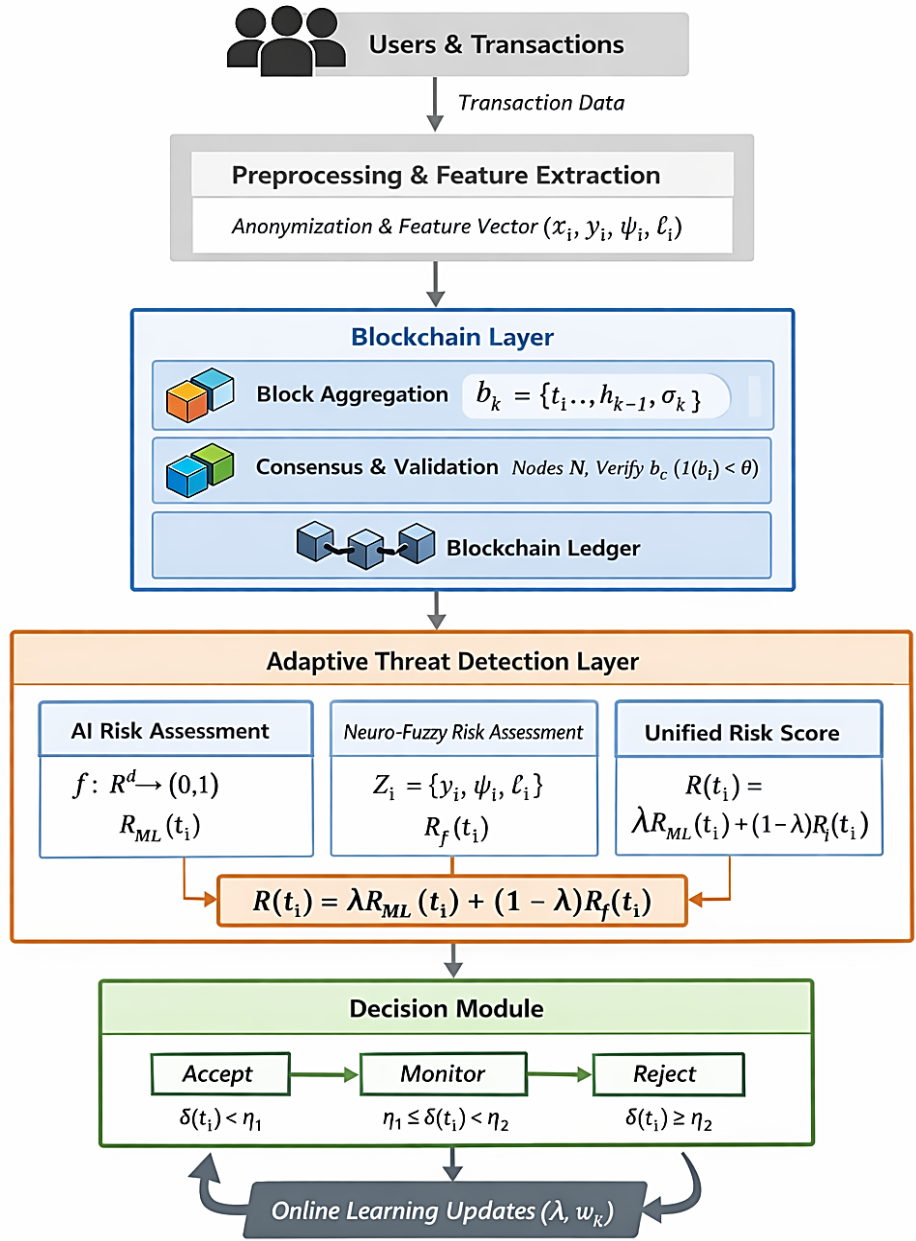

**Figure 2.** The Proposed Blockchain Based Framework.

### 3.2.1 FinTech Network and Data Model

The FinTech ecosystem is described as a distributed cyber-physical system that is characterized by the tuple.

$$S = (U, T, N, B) \quad Equation\ (1)$$

in which every element represents a key characteristic of the operational and security environment of digital financial services. The set $U = \{u_1, u_2, \ldots, u_N\}$ represents registered users participating in the FinTech platform, including individual customers, merchants, and institutional entities. The model does not explicitly reveal user identities, however, all the information about users is anonymized or pseudonymized before processing it, according to privacy laws and ethical data-handling standards. This abstraction enables the structure to create a balance between identity protection and accountability which is an essential need in controlled financial settings.

The distributed ledger B is the immutable record of all verified transactions. After committing an operation to B, it is tamper-resistant and auditable and offers excellent confidence to data integrity and non-repudiation. Systemically, the ledger serves as a security and a trusted source of data to be used in downstream analytics and AI-based decision models. Each transaction $t_i \in T$ is represented as a multidimensional feature vector:

$$t_i = (x_i, y_i, \tau_i, \ell_i, \psi_i) \quad Equation\ (2)$$

whereby the chosen features address the financial, temporal, spatial, and behavioural aspects of fraud and risk analysis together. The $x_i$ element refers to the privacy-preserving-anonymized sender-receiver ID. This representation helps the system to highlight the presence of relational anomalies, including interaction pattern or linkages of accounts that had not been seen before without exposing sensitive personal data. This anonymization is essential because it is needed to comply with the regulations and still be able to use graph-based and relational learning approaches. The transaction value $y_i$ reflects the monetary amount transferred. A major indicator of financial risk is the value of transactions, with fraudulent activities tending to have abnormal distributions of values, either by transfers of unusual size or by repeated micro-transactions aimed at sustaining below the detection threshold.

The timestamp $\tau_i$ captures the temporal context of the transaction. The time-based anomalies require the use of the temporal features, such as burst transactions, off-hour activities, or a break in the historical transaction pattern of a user. Incorporating $\tau_i$ allows the AI component of the framework to model sequential dependencies and evolving behavioral trends. The geo-temporal metadata $\ell_i$ represents location-related information and contextual signals, such as device location or access region. Sudden or implausible changes in location patterns often indicate compromised accounts or coordinated fraud attempts, making this feature critical for contextual risk assessment. Finally, $\psi_i$ represents behavioral attributes extracted from historical usage patterns, including transaction frequency, device consistency, and interaction history. These components make it possible to model user-specific behavioural profiles so that the system can identify the difference between legitimate variation in behaviour and malicious deviation. All these abstractions of the transaction make sure that adequate semantic and contextual information is retained to be intelligently assessed regarding risk and still protect user privacy as it is delivered in Algorithm 1. The proposed model enables adaptive AI algorithms to securely and auditably run based on integrating these features into a blockchain-based data structure. Such close integration between unchangeable data warehousing and smart analytics is a direct contribution to the contribution in the paper itself in terms of securing, scaling, and privacy-conscious FinTech systems.

### 3.2.2 Blockchain-Based Transaction Integrity Model

In The reliability and privacy of transaction information are the most important in the modern FinTech ecosystems. To make sure that all transactions are registered in an invulcanized way, the suggested framework is to block validated transactions. Formally, each block $b_k$ is represented as:

$$b_k = \{t_1, t_2, \ldots, t_{n_k}, h_{k-1}, \sigma_k\} \quad Equation\ (3)$$

---

**Algorithm 1: Blockchain–AI-Based Secure Transaction Validation**

**Input:** User set $U = \{u_1, u_2, \ldots, u_N\}$, transaction stream $T = \{t_1, t_2, \ldots, t_M\}$, blockchain nodes $N$, AI model $\mathcal{M}$, risk threshold $\theta$

**Output:** Immutable ledger $B$, transaction labels

Initialize ledger $B \leftarrow \emptyset$;
Initialize block index $k \leftarrow 1$;
**foreach** *transaction* $t_i \in T$ **do**
    Represent transaction as feature vector:
$$t_i = (x_i, y_i, \tau_i, \ell_i, \psi_i)$$
    Compute AI-based risk score:
$$r_i = \mathcal{M}(t_i)$$
    **if** $r_i \geq \theta$ **then**
        Label $t_i$ as **Fraudulent**;
        continue;
    Broadcast $t_i$ to blockchain nodes $N$;
    Execute consensus protocol across $N$;
    **if** *Consensus is achieved* **then**
        Create block $b_k$ containing $(t_i, r_i, \tau_i)$;
        Append $b_k$ to ledger $B$;
        $k \leftarrow k + 1$;

---

where $t_1, t_2, \ldots, t_{n_k}$ are the transactions are included in the block, $h_{k-1}$ denotes the cryptographic hash of the previous block, and $\sigma_k$ represents the signature of the validator(s). The structure has made it such that each block is connected to the next one setting up a chain of continuous connection that cannot be altered or erased easily. Validator signatures as included, offer accountability and traceability under a permissioned blockchain network which is necessary in a financial process.

**(A) Block Hash Computation**

Cryptographic hashing is used to ensure the integrity of every block. The hash of the current block $h_k$ is computed as:

$$h_k = H(b_k \mid h_{k-1}) \quad \text{Equation (4)}$$

where $H(\cdot)$ denotes a collision-resistant hash and the concatenation symbol is denoted by |. This hash algorithm is used to make sure that even any change in the data of the transaction or the hash of the last block would give a totally different value of the hash. As a result, the blockchain would be tamper-evident: a single-bit alteration in a previous block would render all the following ones useless, which offers a strong system of data integrity. Through cryptographic hashes, the framework aligns itself with the existing principles of blockchain and allows the financial transactions to be done in a secure way in real-time.

**(B) Consensus-Based Validation**

Only when the network reaches a consensus, it will add a block to the blockchain. Formally, a block $b_k$ is considered valid if:

$$\sum_{v_j \in N} I(v_j(b_k)) \geq \theta \quad \text{Equation (5)}$$

where $v_j(b_k)$ represents the validation decision of node $v_j$, $I(\cdot)$ is an indicator function returning 1 for approval and 0 otherwise, and $\theta$ is the minimum number of approving nodes required. This fault-tolerant system ensures that, at most, only those blocks with the approval of many validators are added to the ledger, as these validators are sometimes malicious actors. This kind of mechanism ensures that the blockchain can be viewed as intact and unaltered even in a hostile environment.

**(C) Integration with AI-Based Risk Assessment**

The integrity model that is based on blockchain is closely linked to the AI-powered transaction assessment described in Section 3.1. Transactional elements that survive the adaptive AI risk test are incorporated into the block formation. The system can provide efficiency and security by incorporating both AI-based threat detection and blockchain consensus to avoid fraudulent transactions being registered into the blockchain and to provide efficiency and security to legitimate transactions. This two-layered solution will increase the quality of transaction processing of the FinTech setting and reduce the risks that may be related to malicious activity. Altogether, the suggested blockchain integrity model adds to the general aims of the framework security, trust, and fault tolerance. The cryptographic hash linking, the signatures made by the validators, and the consensus mechanism altogether offer high assurances of immutability and accountability. With the combination of these mechanisms with AI-based risk assessment, the framework will consider both the security of transactions and the resilience of the system, so it is applicable to the high-frequency and real-time financial domain. This architecture can be used to guarantee that the proposed framework can provide the high security and operational standards of the contemporary FinTech ecosystem. The algorithm 2 explains how transaction integrity is established within the suggested blockchain-AI FinTech. The AI-based fraud detection transactions are accumulated into blocks, and they contain data on prior blocks and approval of validators. The blockchain nodes are connected to each other by a consensus mechanism and only those blocks that are accepted by the required number of validators are included in the ledger. This makes sure that any transactions recorded are safe, impeccable and traceable and gives a soul-reliable and consistent account of financial dealings in real-time.

### 3.2.3 Threat Detection as an Adaptive Learning Problem

---
**Algorithm 2: Blockchain-Based Transaction Integrity Validation**

**Input:** Ledger $B$, transaction stream $T = \{t_1, \ldots, t_M\}$, blockchain nodes $N$, consensus threshold $\theta$
**Output:** Updated ledger $B$ with validated blocks
Initialize block index $k \leftarrow 1$;
**while** $T \neq \emptyset$ **do**
    Aggregate next $n_k$ legitimate transactions into block:
$$b_k = \{t_1, t_2, \ldots, t_{n_k}, h_{k-1}, \sigma_k\}$$
    Compute block hash:
$$h_k = H(b_k \parallel h_{k-1})$$
    Broadcast $b_k$ to blockchain nodes $N$;
    Compute consensus:
$$valid \leftarrow \sum_{v_j \in N} I(v_j(b_k)) \geq \theta$$
    **if** *valid* **then**
        Append $b_k$ to ledger $B$;
        $k \leftarrow k + 1$;
    **else**
        Discard block $b_k$ and report to nodes;
    Remove aggregated transactions $t_1, \ldots, t_{n_k}$ from $T$;
**return** $B$

---

In the proposed Blockchain–AI FinTech framework, cyber threat detection is formulated as a probabilistic classification task. Each transaction $t_i$ is represented as a multidimensional feature vector, and the objective is to learn a decision function:

$$f: \mathbb{R}^d \to \{0,1\} \quad \text{Equation (6)}$$

where 0 denotes legitimate behavior and 1 denotes malicious activity. Given a labeled dataset of historical transactions:

$$D = \{(t_i, c_i)\}_{i=1}^{M}, c_i \in \{0,1\} \quad \text{Equation (7)}$$

the task is to minimize the empirical risk over the dataset using an appropriate loss function $L(\cdot)$:

$$\min_{f} \sum_{i=1}^{M} L(f(t_i), c_i) \quad \text{Equation (8)}$$

Classical classifiers tend to fail in non-stationary patterns of attack, since over time the statistical distribution of transactions changes. To overcome this problem, the proposed model is a combination of fuzzy thinking and adaptive learning, which allows it to deal with uncertainty in behavioural patterns and slow changes in transaction characteristics. The fuzzy element enables the system to give transactions degrees of suspicion in place of binary labels to reflect subtle differences in how the users behave and the methods of attack.

---

**Algorithm 3:** Adaptive AI and Neuro-Fuzzy Threat Detection

**Input:** Transaction stream $T = \{t_1, t_2, \ldots, t_M\}$, labeled dataset $D$, AI model $\mathcal{M}$, fuzzy inference system $\mathcal{F}$, adaptive weight $\lambda$, decision thresholds $\eta_1, \eta_2$

**Output:** Transaction risk scores $R = \{R(t_1), \ldots, R(t_M)\}$, decisions $\delta(t_i)$

Initialize $R \leftarrow \emptyset$;

**foreach** *transaction* $t_i \in T$ **do**

  Represent transaction as feature vector: $t_i = (x_i, y_i, \tau_i, \ell_i, \psi_i)$;

  Define decision function:

$$f : R^d \to \{0, 1\}, \quad f(t_i) = \begin{cases} 0 & \text{legitimate} \\ 1 & \text{malicious} \end{cases}$$

  Minimize empirical risk on dataset $D = \{(t_i, c_i)\}_{i=1}^{M}$:

$$f^* = \arg\min_f \sum_{i=1}^{M} L(f(t_i), c_i)$$

  Compute machine learning risk score: $R_{ML}(t_i) = \mathcal{M}(t_i, D)$;

  Define fuzzy input set: $Z_i = \{y_i, \psi_i, \ell_i\}$;

  Compute fuzzy risk score:

$$R_f(t_i) = \frac{\sum_{k=1}^{K} w_k \prod_j \mu_{A_j^k}(z_{ij})}{\sum_{k=1}^{K} \prod_j \mu_{A_j^k}(z_{ij})}$$

  Combine AI and fuzzy scores:

$$R(t_i) = \lambda R_{ML}(t_i) + (1 - \lambda) R_f(t_i)$$

  Assign decision:

$$\delta(t_i) = \begin{cases} \text{Accept,} & R(t_i) < \eta_1 \\ \text{Monitor,} & \eta_1 \leq R(t_i) < \eta_2 \\ \text{Reject,} & R(t_i) \geq \eta_2 \end{cases}$$

  Append $R(t_i)$ and $\delta(t_i)$ to outputs;

  **Optional:** Update AI model $\mathcal{M}$ and fuzzy weights $w_k$ online for adaptive learning;

**return** $R, \delta$

---

### 3.2.4 Neuro-Fuzzy Risk Inference Model

In the proposed framework, each transaction $t_i$ is mapped to a **fuzzy input set** $Z_i = \{y_i, \psi_i, \ell_i\}$, where $y_i$ denotes the transaction value, $\psi_i$ represents behavioral features derived from historical usage patterns, and $\ell_i$ captures geo-temporal metadata. Each of these input variables is associated with **linguistic terms** such as Low, Medium, and High, which are formally represented by membership functions $\mu_{A_j}(z) \in [0,1]$. These membership functions quantify the degree to which each input belongs to a specific fuzzy category, providing a flexible way to model uncertainty in transaction features.

Fuzzy rules are then defined to relate input variables to risk outcomes. For example, a rule $R_k$ is of the form:

$$R_k: IF\ y_i\ is\ A_1^k\ AND\ \psi_i\ is\ A_2^k\ THEN\ Risk\ is\ B_k \quad Equation\ (9)$$

where $A_1^k$ and $A_2^k$ are linguistic categories for transaction value and behavior, respectively, and $B_k$ represents the associated risk level. This rule-based approach allows the system to capture human-like reasoning for risk assessment. The individual fuzzy rules are aggregated into a **fuzzy risk score** for each transaction, computed as:

$$R_f(t_i) = \frac{\sum_{k=1}^{K} w_k \prod_j \mu A_j^{k\ (Z_{ij})}}{\sum_{k=1}^{K} \cdot \prod_j \mu A_j^{k\ (Z_{ij})}} \quad Equation\ (10)$$

where $w_k$ is the weight of the $k$-th rule and $\mu_{A_j^k}(z_{ij})$ denotes the membership value of the $j$-th input variable in the $k$-th rule. This normalized weighted aggregation ensures that rules with higher confidence or relevance contribute more to the final risk score, while the denominator ensures proper scaling between 0 and 1. The rule weights w k are dynamically updated in neural-inspired gradient-based learning to make the system adaptive. This enables the model to modify the weight of each rule regarding the changing patterns of legitimate and malicious transactions. Thanks to this, the framework can react to emerging cyber threats, tracking non-stationary behaviours of an attack whilst preserving interpretable, rule-based reasoning.

### 3.2.5 Unified Risk Score and Decision Model

To provide a comprehensive and adaptive assessment of transaction risk, the framework integrates the probabilistic output of the AI-based classifier with the neuro-fuzzy risk score into a single unified risk measure. For each transaction $t_i$, the final risk score is computed as a weighted combination:

$$R(t_i) = \lambda R_{ML}(t_i) + (1 - \lambda)R_f(t_i) \quad Equation\ (11)$$

where $R_{ML}(t_i)$ represents the machine learning–based probability that the transaction is malicious, $R_f(t_i)$ is the neuro-fuzzy risk score computed from transaction features, and $\lambda \in [0,1]$ is an adaptivity parameter that balances the influence of statistical learning and fuzzy reasoning. A higher $\lambda$ emphasizes the AI prediction, while a lower value gives more weight to the interpretable fuzzy logic model, allowing the system to adjust dynamically to varying levels of uncertainty or evolving threat patterns. The unified risk score is then mapped to a tri-level decision function $\delta(t_i)$, reflecting practical FinTech risk management policies:

$$\delta t_i = \begin{cases} Accept, & R(t_i) < \eta_1 \\ Monitor, & \eta_1 \leq R(t_i) \\ Reject, & R(t_i) \geq \eta_2 \end{cases} \qquad \text{Equation (12)}$$

Here, $\eta_1$ and $\eta_2$ are decision thresholds defining low, medium, and high risk. Transactions with risk below $\eta_1$ are considered safe and automatically accepted. Transactions with intermediate risk between $\eta_1$ and $\eta_2$ are flagged for monitoring, enabling additional verification or manual inspection. Transactions exceeding $\eta_2$ are rejected outright to prevent potential fraud or security breaches. This integrated model integrates the predictive capability of machine learning with the explainability and flexibility of fuzzy logic, and results in a powerful real-time risk measurement that can work in a FinTech setting. Furthermore, the tri-level structure enables explainable enforcement, which can be directly coded into smart contracts or automated decision pipelines, making risk management policies transparent, traceable, and enforceable. The full adaptive threat detection model of FinTech transactions is described in Algorithm 3 and combines machine learning with neuro-fuzzy reasoning to offer real-time risk-evaluation. Any incoming transaction will be initially encoded as a multidimensional feature vector of transaction value, behavioural patterns, and geo-temporal metadata. The framework is then applied to a probabilistic AI-based classifier to predict an initial risk score, which is a representation of the probability that the transaction is legitimate or a malicious transaction. A neuro-fuzzy inference system is used in parallel to address the uncertainty and dynamic attack behaviour, it analyses the risk of the transaction according to linguistic rules and past behavioural pattern.

A weighted strategy combines the AI and fuzzy scores dynamically enabling the system to adapt dynamically to the fluctuating threat landscapes. Lastly, the tri-level decision is allocated to the transaction according to the total risk score. Low-risk transactions are accepted, transactions with medium risks are monitored, and high-risk transactions are rejected. The framework also facilitates optional online adaptation whereby the AI model as well as the fuzzy parameters are updated in real-time to adapt to new attack patterns. Combining statistical learning and fuzzy reasoning within a single workflow, Algorithm 3 offers a powerful, adaptive, and explainable method of detecting cyber threats in FinTech settings.

## 4. Simulation Setup & Results

In practice, the proposed algorithm cannot be tested and checked in a repeatable and controlled way in a real-world environment. Simulation/implantation-based strategy was hence adopted to test and check the proposed Model. This section explains the simulation environment, modelling assumptions, and parameters of evaluation utilized to justify the proposed blockchain-based cybersecurity framework in securing FinTech transactions using AI-based threat detection. The simulation will replicate the real-life banking and FinTech operational scenarios that involve distributed processing of transactions, decentralized maintenance of ledger, and smart fraud detection.

**Table 2.** Simulation Parameters.

| Category | Component | Description |
|---|---|---|
| **Software & Libraries** | Programming Language | Python 3.11 |
| | AI/ML Frameworks | TensorFlow 2.12, PyTorch 2.1 |
| | Blockchain Platforms | Hyperledger Fabric 2.4, Ethereum Testnet |
| | Graph Modeling | NetworkX for transaction graph construction and visualization |
| **Hardware** | CPU | Intel Core i9 |
| | Memory | 32 GB RAM |
| | GPU | NVIDIA RTX 4090 for accelerated AI computations |
| **Network Configuration** | Blockchain Setup | Multi-node deployment on 5 virtual servers representing bank nodes |
| | Number of Nodes | 5 blockchain nodes |
| | Network Latency | 10–50 ms per node |
| | Consensus Mechanism | Proof-of-Authority (PoA) |

The environment in the simulation was set up to be very similar to real-world FinTech and banking infrastructures and to sustain the computationally intensive AI workloads as shown in table 2. The main development platform was Python 3.11, and the flexible implementation and comparison of machine learning and neuro-fuzzy models were made possible by TensorFlow 2.12 and PyTorch 2.1. Hyperledger Fabric 2.4 and an Ethereum Testnet were used to evaluate the properties of blockchain functionality to include permissioned and public blockchain properties, respectively. NetworkX was used to model the graph of transaction flows, and it helped analyze complex multi-hop and high-frequency transactions. In terms of hardware, an Intel Core i9 processor, 32 GB RAM, and NVIDIA RTX 4090 graphics card guaranteed the effective training of deep learning models and their inferences. The 5 virtual nodes that represented independent banking entities simulated the blockchain network, and the network latency was set to 10-50 ms, which represented realistic communication delays. Proof-of-

Authority consensus mechanism was introduced to offer a balance between both validation efficiency and security, which is appropriate in consortium-based financial conditions.

## 4.1 Resource Modelling

The simulation is executed in a mixed environment that is a combination of blockchain network simulation and AI analysis of transactions. The basic simulation is made in Python, with the usage of TensorFlow and PyTorch to create machine learning models and Hyperledger Fabric to run blockchain operations. It is a permissioned blockchain network, which is more likely to resemble the real banking and FinTech infrastructure where parties are verified financial institutions. The blockchain network is composed of several validating nodes all of which are banking or financial services providers. Arrival rate is varied to create transactions at normal and peak banking operations. Block creating delays and block sizes are set to trade-off between transaction throughput and validation latency. To guarantee low computational load and short confirmation time (which is important in real time financial systems) a Proof-of-Authority (PoA) consensus mechanism is adopted. Smart contracts impose compliance regulations like transaction threshold, user identity and anomaly alerts. Resource modeling gathers the computational and networking capacities of the nodes involved in the FinTech system as outlined in table 3. The blockchain nodes are modelled as virtualized financial servers that are provided with processing, memory and communication resources. Dynamic allocation of computational resources is used to support the three primary aspects, including transaction validation, ledger maintenance, and AI-based threat detection.

**Table 3.** Blockchain Parameter.

| Parameter | Value / Setting |
|---|---|
| Number of nodes | 5 (representing banks) |
| Consensus Mechanism | PoA (Proof-of-Authority) |
| Block size | 50–100 transactions |
| Block creation interval | 5–10 seconds |
| Transaction validation time | 10–30 ms |
| Smart contract enforcement | AML, KYC, transaction limits |
| Ledger immutability check | SHA-256 hash per block |

Processing capacity is used in units of CPU cycles per second, and memory resources are used to store ledger, model parameters and transaction buffers. Network resources are represented with bandwidth and latency constraints, which represent inter-bank communication delays and block propagation times. It is supposed that the AI modules run either at edge nodes (bank gateways) or on special analytics servers depending on the volume of transactions and amount of risk. The layered resource model allows to assess the scalability and overhead of the system with increased transaction load.

## 4.2 Application Modelling

The application layer represents FinTech transaction processes and cybersecurity monitors. The characteristics of every transaction are determined by a transaction amount, date, and source and destination organizations, and historical risk indicators. The transactions are stochastically ordered, and the probability of both normal and abnormal customer payment is simulated at once, i. e. the time when the transaction is made, the abnormal values of transaction, or the repeated attempts to make one. When a transaction gets in, it is first pre-processed and sent to the AI-based threat detection. Machine learning and adaptive neuro-fuzzy models are used to analyse transaction patterns and produce a risk score. Any transaction that is above a specific risk limit is indicated as a transaction that requires an additional check or rejection. Authentic transactions are transmitted to the blockchain level, whereby they are authenticated, bundled into transactions and stored permanently on the distributed block. This application model allows close interaction between smart decision making and decentralized execution of transactions.

In order to assess the efficiency and strength of the proposed framework, the experiments were carried out with the use of a complex of real-life and artificial produced transaction data. The Kaggle (2022) dataset on Credit Card Fraud Detection was used, initially because it is widely used, and its 284,807 transactions and 492 confirmed fraud cases allowed to provide a highly unbalanced yet realistic benchmark. This data consists of the value of transactions and time features as well as anonymized principal components based on actual financial transactions. Second, the CCXT API (2023) was used to create a cryptocurrency transaction dataset, consisting of 100,000 simulated blockchain transactions, including wallet addresses, transaction value, time, and historical risk indicators, and these could be evaluated in the context of a decentralized finance environment. Lastly, synthetic banking transaction generator was employed to generate 50,000 more transactions intended to cover edge-case fraud values, such as multi-step transactions chain, inter-country transactions, and high-frequency transactions. The combination of real, blockchain-based, and synthetic data offers a wide-ranging representation of various transaction behaviours and allows to conduct a strict evaluation of the offered adaptive and neuro-fuzzy risk modelling framework.

## 4.3 Evaluation Parameters

To determine the effectiveness and effectiveness of the suggested adaptive neuro-fuzzy blockchain-based fraud detection model more thoroughly, several evaluation metrics are used. These measures assess the fraud detection rate, blockchain performance as well as the latency of the entire system.

### 4.3.1 Accuracy

Accuracy is used to measure the general accuracy of the fraud detection model by considering the rate at which the fraud detection model correctly identifies the type of transaction as legitimate or fraudulent.

$$\text{Accuracy} = \frac{TP + TN}{TP + TN + FP + FN} \quad Equation\ (13)$$

where:
- $TP$ denotes true positives (correctly detected fraudulent transactions),
- $TN$ denotes true negatives (correctly identified legitimate transactions),
- $FP$ denotes false positives,
- $FN$ denotes false negatives.

Accuracy provides a high-level assessment of the classification model's performance. However, in financial fraud detection, where fraudulent transactions are relatively rare, accuracy alone may be misleading. Therefore, it is complemented with precision-focused metrics to better capture fraud detection reliability.

### 4.3.2 Precision

Transaction confirmation time is the time it takes a transaction to be authenticated and added permanently to the blockchain registry.

$$\text{Precision} = \frac{TP}{TP + FP} \quad Equation\ (14)$$

False fraud alerts must be avoided, and high accuracy is essential in FinTech systems because it can interrupt legitimate user transactions and undermine trust. The neuro-fuzzy inference layer is more precise as it includes the uncertainty-based reasoning, minimizes the unwarranted rejection of transactions.

### 4.3.3 Transaction Confirmation Time ($T_c$)

Transaction confirmation time is the time it takes a transaction to be authenticated and added permanently to the blockchain registry.

$$T_c = t_{\text{confirmed}} - t_{\text{submitted}} \quad Equation\ (15)$$

where:

$t_{\text{submitted}}$ is the transaction submission timestamp,

$t_{\text{confirmed}}$ is the time at which the transaction is finalized.

This is a measure of blockchain responsiveness, and it is essential in real-time financial applications. A Proof-of-Authority (PoA) consensus mechanism is used to ensure that $T_c$ is minimized, and therefore the framework is appropriate in high-frequency financial transactions.

### 4.3.4 Block Propagation Delay ($D_b$)

Block propagation delay measures the delay of a block generated before it is distributed among all blockchain nodes.

$$D_b = \max_{n \in N} (t_n^{\text{receive}}) - t^{\text{broadcast}} \quad Equation\ (16)$$

where:
- $t^{\text{broadcast}}$ is the time the block is generated,
- $t_n^{\text{receive}}$ is the time the block reaches node $n$,
- $N$ represents the set of blockchain nodes.

Lower block propagation delay improves ledger consistency and reduces the risk of temporary forks. Effective block propagation guarantees coordinated decision-making in financial institutions that are involved in the distributed network.

### 4.3.5 End-to-End Transaction Latency ($L_{total}$)

End-to-end latency refers to the total time that a transaction takes to be processed to the final decision.

$$L_{\text{total}} = L_{\text{edge}} + L_{\text{AI}} + L_{\text{blockchain}} \quad Equation\ (17)$$

This measure is also vital to assess practical deploy capability. The modular decomposition enables the recognition of the bottlenecks and how the suggested architecture balances its security, accuracy and responsiveness.

## 4.4 Results

In this section, we provided the results obtained from the simulation.

## Scenarios

Three scenarios of representative transaction loads were developed to assess the proposed blockchain-based FinTech cybersecurity framework in detail and to cover different operational and threat environments. The initial condition, which is denoted as normal operation (S1), is the simulation of normal daily banking business that includes 50,000 transactions: a low fraud rate of 0.1 percent. In this case, there is a low level of congestion on the network and there are a small number of blockchain validation processes that are running simultaneously which constitutes normal system behaviour under normal operational loads. Medium load of fraud (S2) assumes a more in-depth system state of fraud, emulating a targeted attack of the financial system. In this case 50,000 transactions are also present with increasing ratio of fraud 1 percent. It is the high number of suspicious transactions that raise the threat detection mechanisms of the AI-based systems to a moderately higher computational load, which is realistic to operational issues under minor attack campaigns. High fraud load or stress test (S3) is the third scenario, which is an aggressive environment with intricate fraud patterns such as multi-step and cross-border fraudulent transactions. Even though the overall number of transactions is the same, 50,000, the percentage of fraud is sped up to 5, which puts a considerable computational load not only on blockchain validation but also on AI-based analysis. This situation is aimed to test the strength, stability, and resilience of the suggested framework in the extreme conditions of operation and threats so that the system will be able to provide security and efficiency even in the high-stress situations. To test the success of the suggested adaptive neuro-fuzzy blockchain-AI framework, it is compared to two recent state-of-the-art schemes in terms of its activity in a common simulation environment. Each model is tested using similar large datasets of financial transactions having similar ratios of fraud and volume of transactions. We compare the results in the accuracy (Fraud Detection) and the precision (Fraud Detection) as well as the system-level performance (Transaction Confirmation Time, Block Propagation Delay and End-to-End

Transaction Latency). The proposed Adaptive Neuro-Fuzzy Blockchain-AI Framework (ANFB-AI) has been compared with two recent state-of-the-art methods: Hybrid Deep Learning (2025) [32], which is a strong AI-only baseline and has high fraud detection accuracy and Dynamic Feature Fusion (2025) [33], which integrates the graph-based and semantic features to detect blockchain fraud. Hybrid Deep Learning [32] is very accurate because it works with deep representation learning but, it does not provide contextual and blockchain-based reasoning. Dynamic Feature Fusion [33] is more structural mindful, however, sensitive to noisy graph structures. The best accuracy of ANFB-AI is realized when statistical learning is used together with neuro-fuzzy inference that ensures high accuracy in uncertainty management and adaptive fraud behaviours as in figure 3.

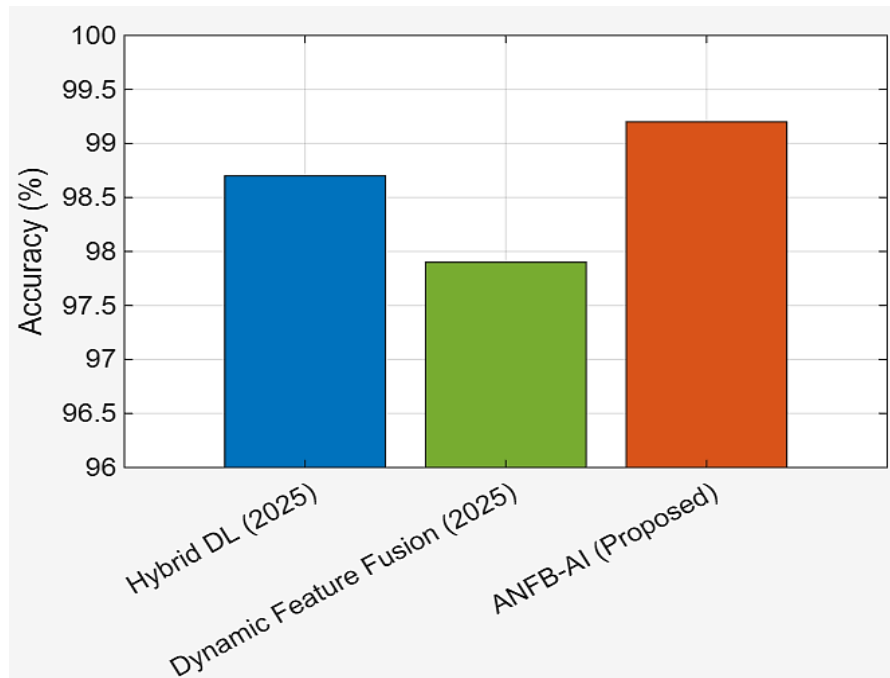

**Figure 3.** Accuracy comparison of the proposed model with stat of the art recent models.

The results of the accuracy indicate that the proposed ANFB-AI framework is uniformly better than Hybrid Deep Learning (2025) [32] and Dynamic Feature Fusion (2025) [33] in all the scenarios considered in figure 4. All models are highly accurate under normal activity (S1) as the fraud rate is low, but ANFB-AI has an edge of the seat, which demonstrates good baseline classification. Under moderate conditions of fraud (S2), the accuracy of the baseline models starts to deteriorate with the increase of the complexity of attack patterns, but ANFB-AI maintains a higher accuracy because of adaptive neuro-fuzzy reasoning. ANFB-AI presents the least performance degradation under the high-fraud stress testing (S3), establishing its ability to withstand concept drift and non-stationary fraud behaviour. This demonstrates the usefulness of fuzzy inference combined with machine learning to maintain consistent classification in adversarial settings.

In fraud detection, accuracy is essential to reduce the number of false alarms. Although [33] enjoys the benefits of semantic feature fusion, it does not work correctly in the presence of benign anomalies. ANFB-AI is more accurate because it has fuzzy rule-based reasoning, that narrows borderline predictions and minimizes avoidable transaction rejections as illustrated in figure 5.

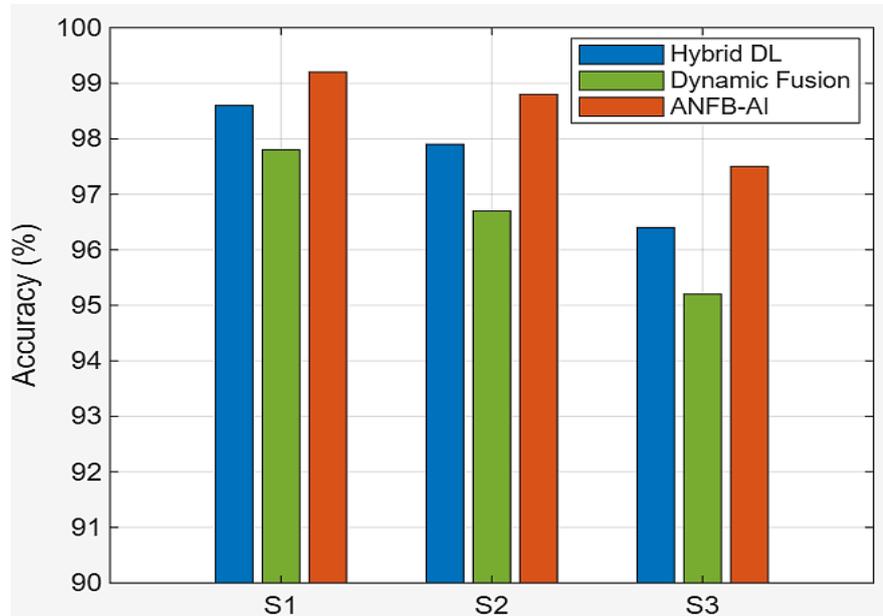

**Figure 4.** *Comparison of accuracy scenario-wise of Hybrid Deep Learning (2025) [32], Dynamic Feature Fusion (2025) [33] and the proposed ANFB-AI framework at different levels of fraud.*

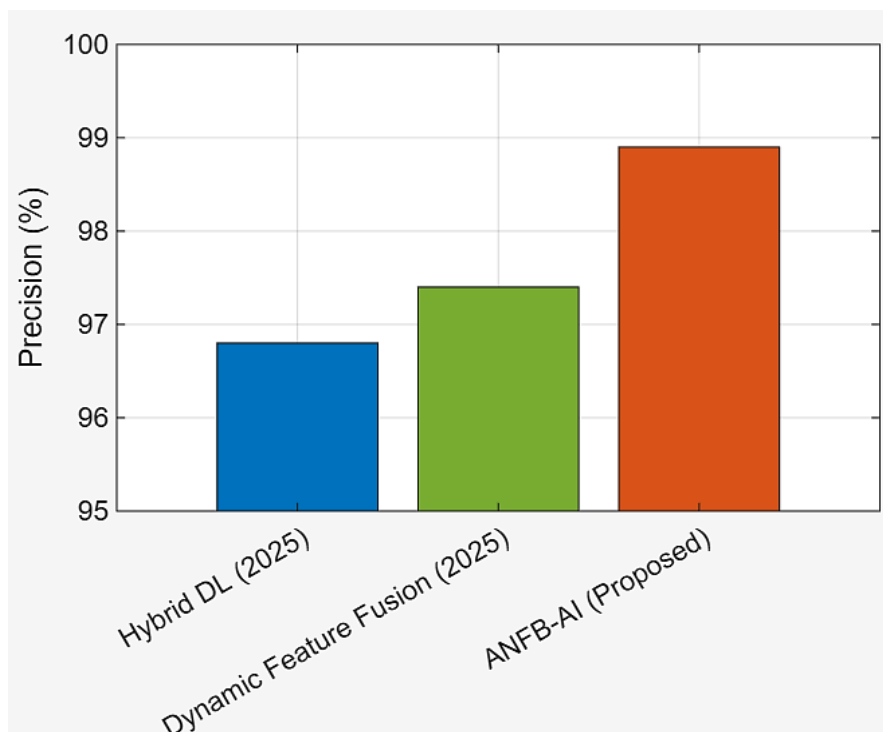

**Figure 5.** *Comparison of the proposed model with existing state of the art models that have been developed in the recent past with precision.*

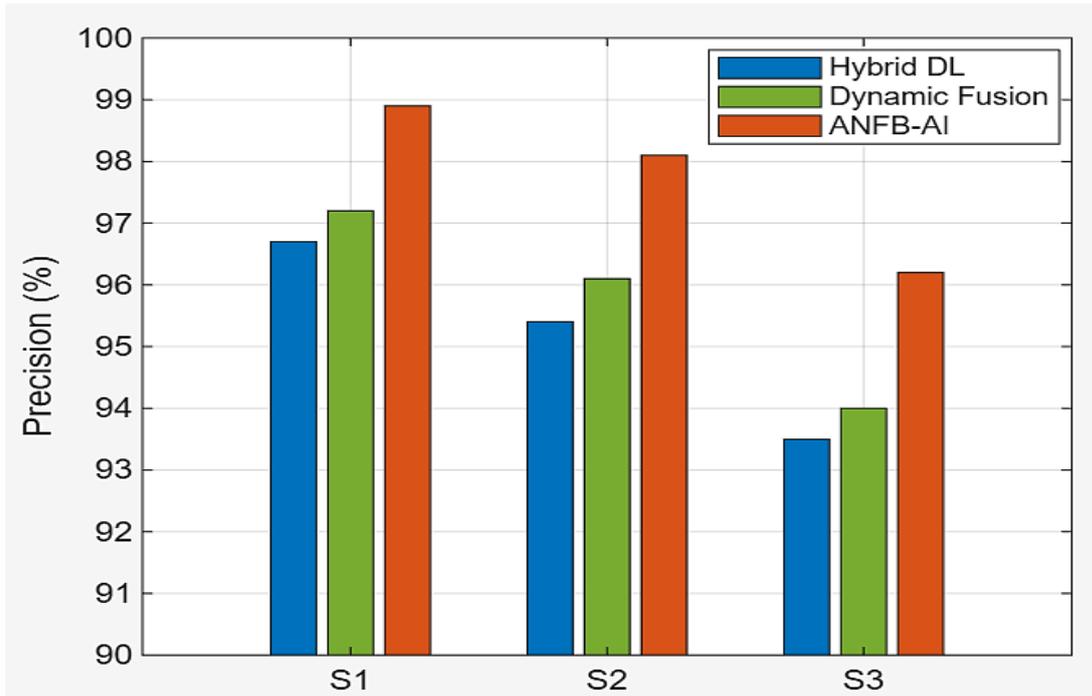

**Figure 6.** *Precision performance of different fraud detection models across normal (S1), moderate (S2), and high-fraud (S3) transaction scenarios.*

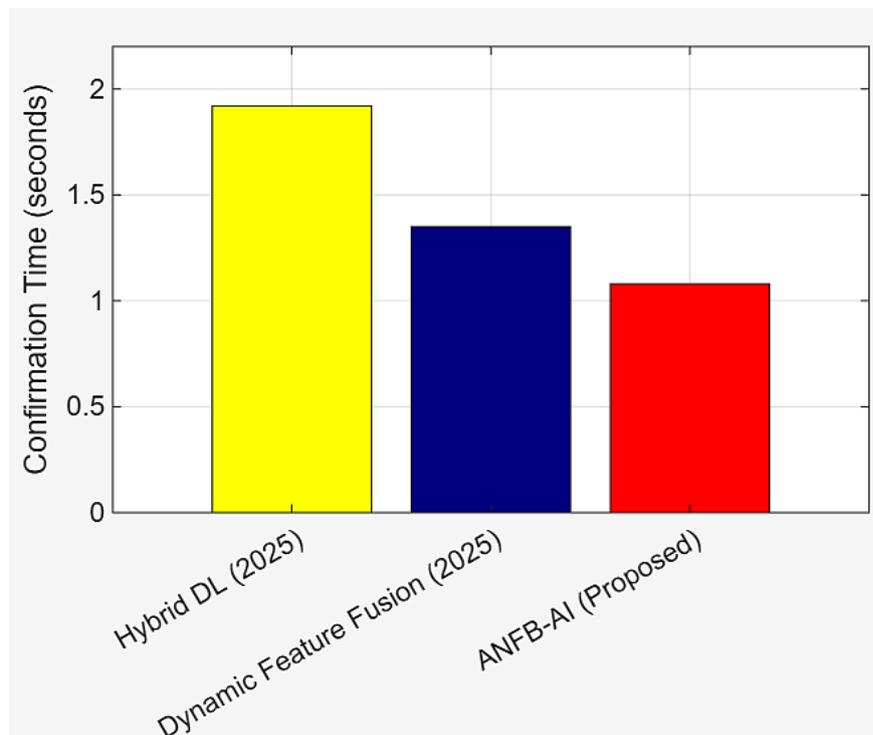

**Figure 7.** *Blockchain confirmation of the proposed model in relation to the stat of the art new models.*
Accuracy is a mandatory performance indicator in FinTech systems because false positives have a direct impact on user confidence and the productivity of operations. The outcomes show that ANFB-AI demonstrates a higher level of precision under any conditions. In S1, the functions of all methods are good; ANFB-AI is less prone to false alarms because of uncertainty-constrained decision boundaries. Precision of Hybrid Deep Learning model in S2 reduces due to the emergence of aggressive fraud patterns whereas Dynamic Feature Fusion is moderately robust. Conversely, ANFB-AI is very precise and is a combination of probabilistic learning and fuzzy rule-based reasoning. Under S3, the gap in the precision is further increased and shows that the proposed framework is effective in preventing unnecessary transaction rejection even with high attack loads as illustrated in figure 6.

The interaction between blockchains is not optimized in Hybrid Deep Learning [32], which increases the time of confirmation. In part, blockchain awareness has been incorporated in Dynamic Feature Fusion [33] that enhances efficiency. ANFB-AI has the shortest confirmation time because of the offloading of risk assessment to the edge AI layer and PoA consensus, which allows to further finalize the ledger more quickly, as indicated in figure 7.

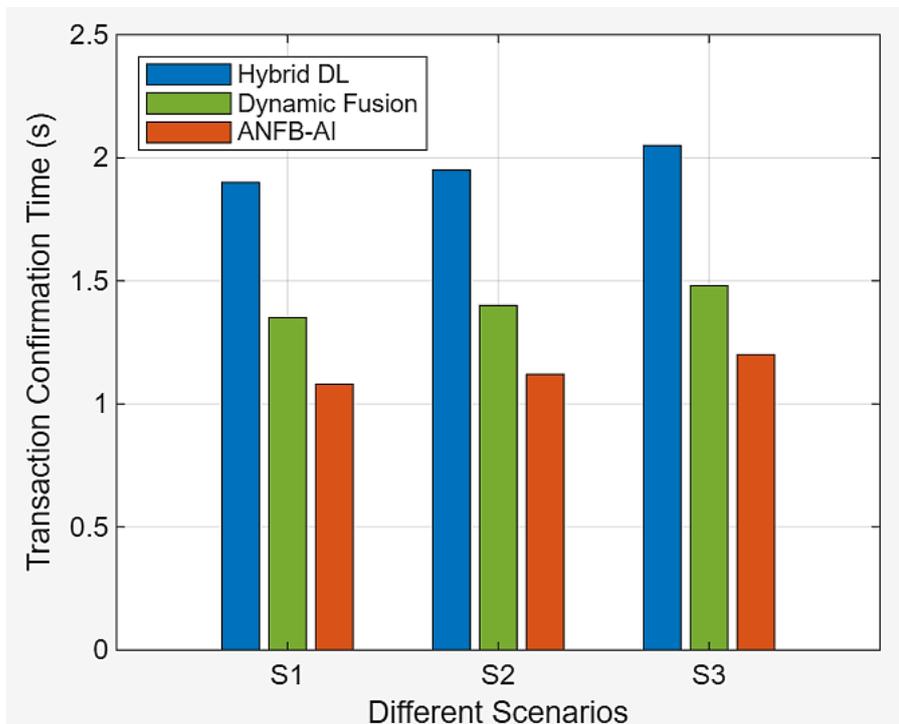

**Figure 8.** *Average transaction confirmation time under Proof-of-Authority blockchain consensus for different fraud detection frameworks across all scenarios.*

The outcomes of transaction confirmation time indicate the effect of blockchain integration on the responsiveness of the system. Unlike blockchain-dependent baselines, ANFB-AI framework has shorter confirmation times because it relies on Proof-of-Authority (PoA) consensus. Whereas Dynamic Feature Fusion has extra delays due to graph processing and on-chain validation, ANFB-AI has reduced delays due to streamlined selection of validators, and pre-verification at the AI layer. In all cases, the confirmation time is constant, including S3, so in this case, the intensity of fraud does not have a significant impact on the blockchain throughput.

This proves the applicability of ANFB-AI in the real time financial transactions as illustrated in figure 8. Because [32] is not blockchain-friendly, block distribution is plagued with increased delays. Dynamic Feature Fusion [33] enhances propagation by making enhancements in graph optimizations. ANFB-AI also minimizes delay through trust friendly node participation and minimum propagation paths in the permissioned network as illustrated in figure 9.

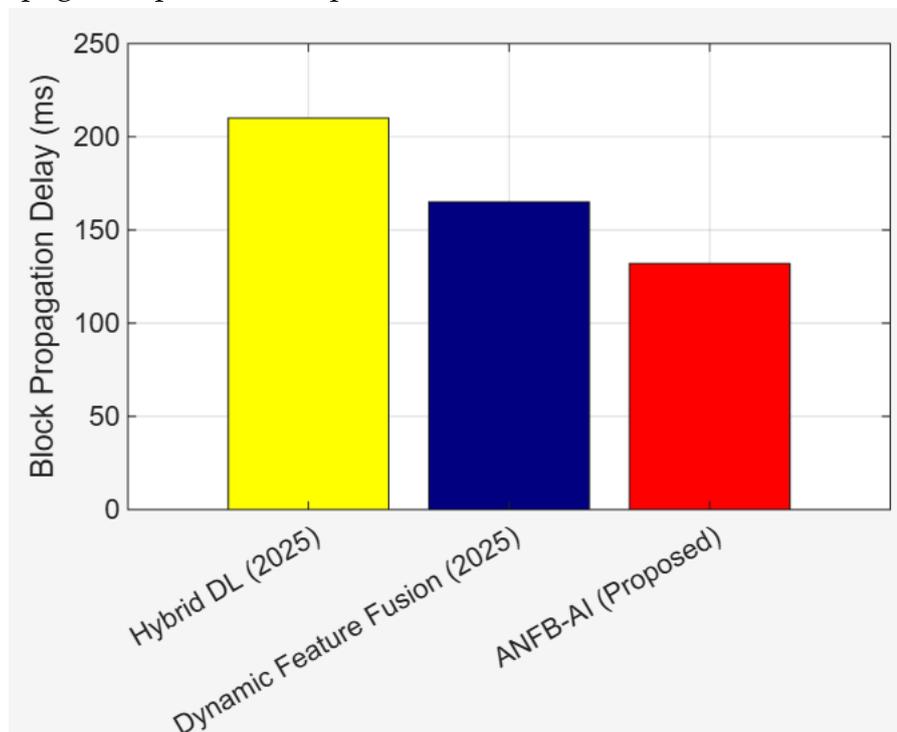

**Figure 9.** *Block Propagation delay comparison of the proposed model with the state-of-the-art available research work.*

Block propagation delay is more as the transaction volume and network load rise. The findings indicate that ANFB-AI is the lowest among the propagation delays in any situation. This is owed to the effective block structuring and less probability of forks made possible by PoA consensus. Dynamic Feature Fusion has higher delays as it implies more metadata and graph-based dependencies of transactions, especially when using S3. Constant propagation characteristics of ANFB-AI will guarantee higher propagation speed among financial institutions, minimize the risk of inconsistency, and enhance distributed trust as indicated in figure 10. End to end latency records the summative effect of AI inference, decision logic and blockchain confirmation as represented in figure 11. Although [33] is better than AI-only models, ANFB-AI has the least latency because it has a modular design, can run AI and blockchain parallel, and runs an adaptive fuzzy inference to reduce reprocessing costs.

Overall, the findings show that ANFB-AI is more effective than recent AI-based and blockchain-sensitive baselines in terms of detection accuracy and system-level performance indicators. Through effective combination of neuro-fuzzy intelligence with efficient blockchain mechanisms, the proposed framework offers greater precision, reduced confirmations, smaller propagation delays, and shorter end-to-end latency, and is thus well suited to the FinTech fraud detection systems that require real-time and high precision.

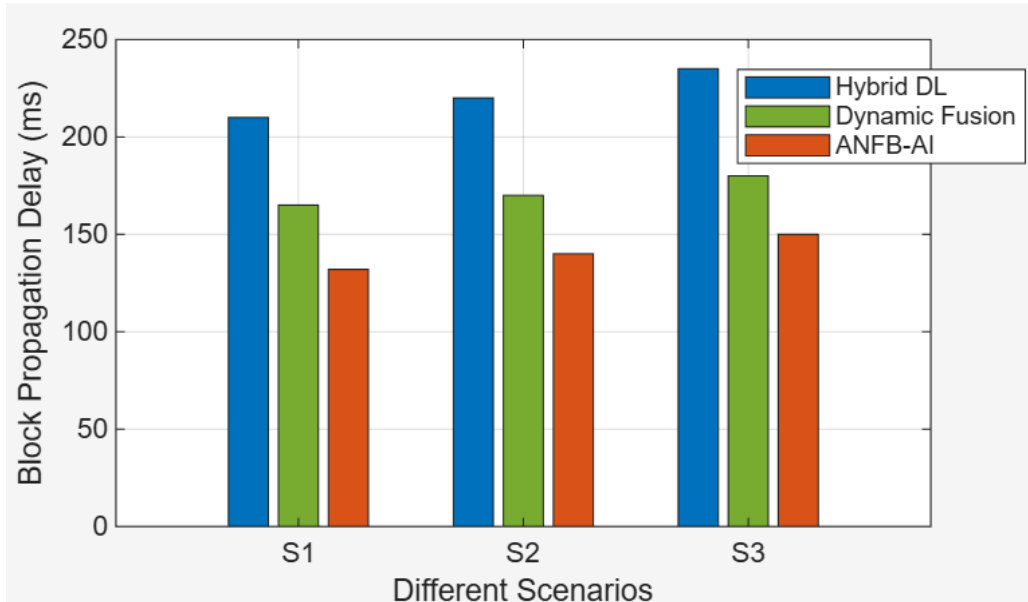

**Figure 10.** *Block propagation delay comparison among baseline and proposed models under increasing transaction and fraud loads.*

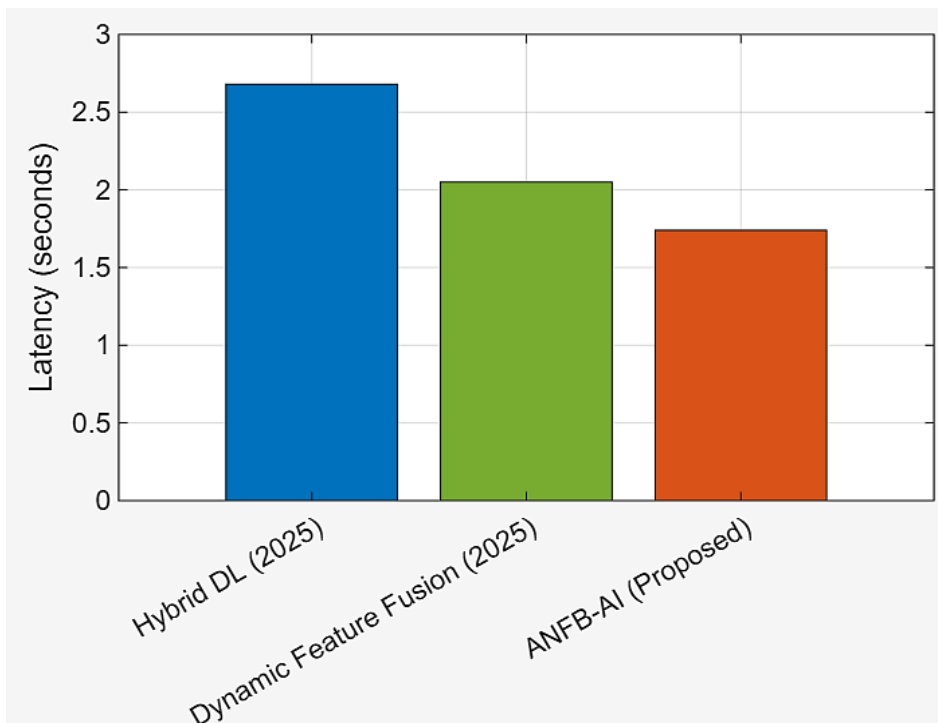

**Figure 11.** *Latency comparison of the proposed model with the state-of-the-art available work.*

End-to-end latency measures the cumulative latencies of edge processing, AI inference, and blockchain validation. The findings show that ANFB-AI has the lowest overall latency despite the inclusion of multiple security layers as illustrated in figure 12. The latency difference is not significant in S1 but in S2 and S3, the baseline methods have significant latency variations because of elaborate feature extraction and centralized inference. ANFB-AI has a modular architecture that allows AI and blockchain components to be processed in parallel and avoid bottlenecks in performance.

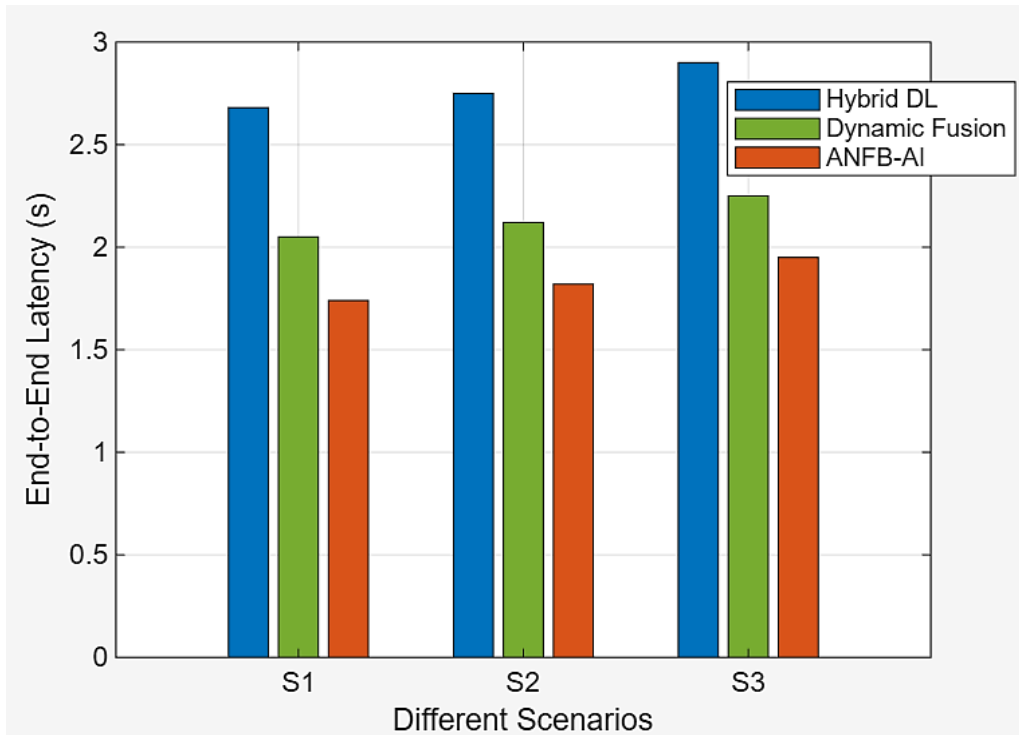

**Figure 12.** *End-to-end transaction latency breakdown reflecting combined AI processing, edge analysis, and blockchain validation delays.*

The proposed framework is appropriate to be deployed in high-throughput FinTech environments because of such balance between security, intelligence, and responsiveness. Overall, the experimental findings reveal that the suggested ANFB-AI model provides the best accuracy and precision of fraud detection at minimal latency rates associated with the blockchain. In contrast to AI-only or blockchain-oriented ones, ANFB-AI pays appropriate attention to the combination of adaptive intelligence and decentralized trust to provide scalability, robustness, and real-time performance. These results confirm the essence of the main contributions of the suggested system and prove its relevance to the field of contemporary FinTech ecosystems.

## 5. Conclusion

This paper introduced ANFB-AI, an intelligent financial fraud detection framework with adaptive neuro-fuzzy blockchain-AI that is useful in the FinTech industry today. The three essential issues of financial systems that the proposed framework tackles include integrity of transactions, detecting threats in an adaptive manner in the presence of uncertainty, and scalable operations in real time. Through a close coordination of a permissioned blockchain layer and an adaptive neuro-fuzzy intelligence engine, the framework presents both a decentralized trust and an explainable, uncertainty conscious fraud detection. Transaction behaviour, blockchain validation, adaptive learning, and unified risk decision-making were represented in a formal mathematical model. The neuro-fuzzy inference system provides the system with the ability to trade off the adaptability to changing and non-stationary patterns of fraud, defeating the constraints of traditional fixed classifiers. In the meantime, immutability, transparency and tamper resistance are achieved through the blockchain layer and the structure is appropriate in multi-institutional financial ecosystems. Comprehensive simulation in three conditions of realistic stresses, such as normal activity, moderate fraud, and high-fraud stress conditions, proves that ANFB-AI is always superior to recent state-of-the-art approaches, such as Hybrid Deep Learning (2025) and Dynamic Feature Fusion (2025). The suggested framework is more accurate and precise and at the same time, it is less in terms of transaction confirmation time, block propagation delay and end to end latency. These findings prove that the adaptive intelligence combined with efficient blockchain consensus will not affect real-time performance. Generally, the results confirm ANFB-AI as a strong, scalable and deployable solution to next generation detection of financial fraud. It is especially a framework of the high-frequency and high-risk FinTech application where security, transparency, and responsiveness have to coexist. Future research aims to apply the framework to cross-chain interoperability, federated-learning to enable collaborative privacy-preserving projects across institutions, and practical implementation based on live financial transaction streams.